\documentclass[3p,times,twocolumn]{elsarticle}
\usepackage{ecrc}
\volume{00}
\firstpage{1}
\journalname{Nuclear Physics B Proceedings Supplement}
\runauth{T. Lappi}
\jid{nppp}
\jnltitlelogo{Nuclear Physics B Proceedings Supplement}
\usepackage{amssymb}
\usepackage[figuresright]{rotating}
\usepackage{tikz}

\usepackage{xcolor}
\definecolor{lcolor}{rgb}{0.5,0,0}
\definecolor{citcolor}{rgb}{0,0.3,0.0}
\usepackage[breaklinks,colorlinks,urlcolor=blue,citecolor=citcolor,linkcolor=lcolor]{hyperref}

\newcommand{\ud}{\, \mathrm{d}}
\newcommand{\nc}{{N_\mathrm{c}}}

\newcommand{\as}{\alpha_{\mathrm{s}}}
\newcommand{\lqcd}{\Lambda_{\mathrm{QCD}}}
\newcommand{\tr}{\, \mathrm{Tr} \, }

\newcommand{\ncoll}{N_\mathrm{coll}}
\newcommand{\npart}{N_\mathrm{part}}

\newcommand{\ptt}{{p_T}}
\newcommand{\ktt}{{k_T}}
\newcommand{\kt}{{\mathbf{k}_T}}

\newcommand{\bt}{{\mathbf{b}_T}}
\newcommand{\pt}{{\mathbf{p}_T}}
\newcommand{\xt}{{\mathbf{x}_T}}
\newcommand{\yt}{{\mathbf{y}_T}}
\newcommand{\rt}{{\mathbf{r}_T}}

\newcommand{\qs}{Q_{\mathrm{s}}}
\newcommand{\fig}{Fig.~}

\begin{document}

\begin{frontmatter}
\dochead{}
\title{Initial state in heavy ion collisions}

\author{T. Lappi}
\address{
Department of Physics, %
 P.O. Box 35, 40014 University of Jyv\"askyl\"a, Finland
and \\
Helsinki Institute of Physics, P.O. Box 64, 00014 University of Helsinki,
Finland
}

\begin{abstract}
We briefly review advances in understanding the initial stages of a heavy 
ion collision. In particular the focus is on  moving from 
parametrizing the initial state to calculating its properties
from QCD, consistently with the description of hard probes 
and dilute-dense scattering experiments.
Modeling the event-by-event fluctuating
nuclear geometry in initial state calculations
has significantly improved in recent years.
We also discuss prospects of directly 
seeing effects of particle correlations created in the initial 
state in the experimental observables.
\end{abstract}


\end{frontmatter}

\section{Introduction}

For a practitioner of hydrodynamic modeling of heavy ion collisions,
the question of initial conditions often boils down to an 
acronym soup of models. One  tries out a suitable subset of
these (``Glauber'', KLN, [mc]KLN, mcrcBK, EPOS, EKRT, IPglasma etc.)
as an input to a fluid dynamics calculation and compares to experimental
data. The purpose here is very different, namely to discuss the physics
ingredients  in these calculations and concentrate more
on their common aspects them than on the differences.
This is done with the caveat that we are concentrating exclusively
on a weak coupling partonic description of the
degrees of freedom involved. We will start with a very brief introduction 
to the physics picture in the weak coupling approach. 
We then discuss dilute-dense
control experiments that can be used to more directly probe the 
degrees of freedom in the initial state of a heavy ion 
collision. The most recent dynamical
initial state models will then be discussed in comparison to 
 Monte Carlo Glauber parametrizations. Finally we will describe recent 
calculations of long range correlations that, if one is able to discriminate
between effects of collective flow and those originating from the initial 
state, could provide a direct access to the latter.

\section{Initial state at weak coupling}

\begin{figure}
\centerline{\includegraphics[height=4cm]{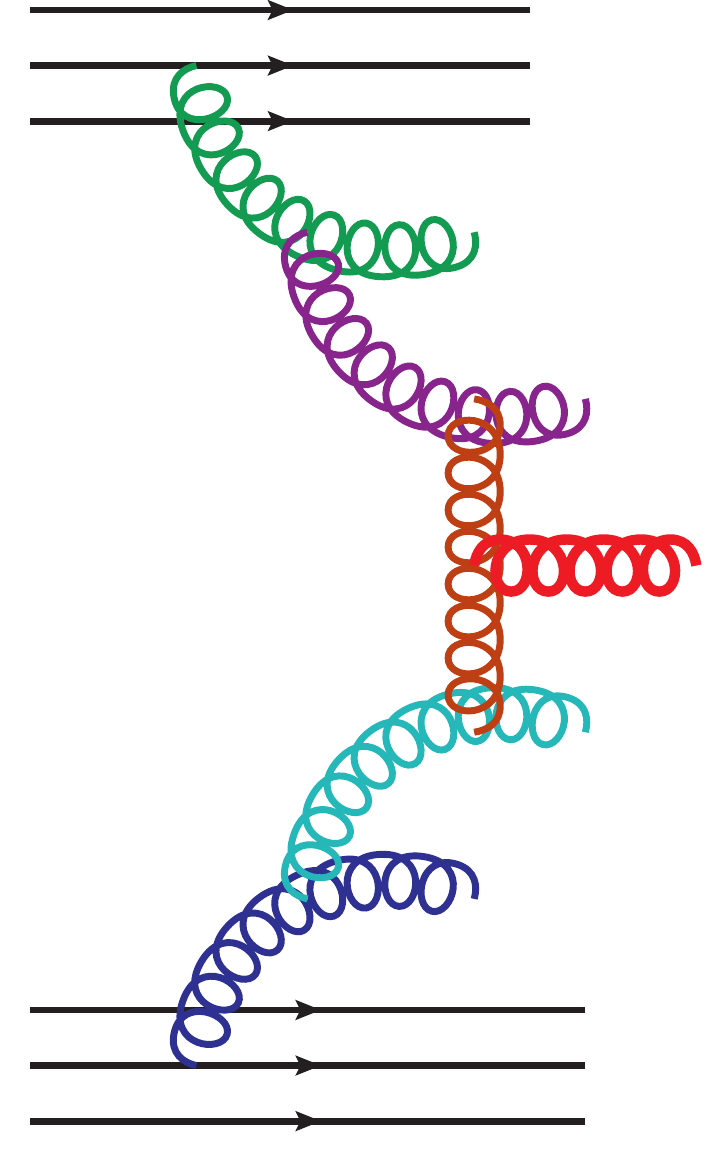}
\begin{tikzpicture}[overlay]
\draw[line width=2pt,<->](-3cm,0.2cm) -- (-3cm,3.8cm);
\node[anchor=west]at (-3,2) {$y\sim \ln \sqrt{s}$};
\end{tikzpicture}
}
\caption{Gluon cascade leading to particle production in the 
central rapidity region.}
\label{fig:casc}
\end{figure}
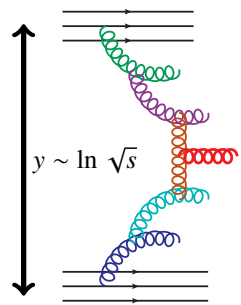

The weak coupling picture of 
particle production at central rapidities in very high energy
hadronic collisions is depicted in \fig\ref{fig:casc}. The gluons
that are mainly responsible for particle production result from 
a cascade of multiple splittings from the initial valence quarks.
The emission probability for one splitting is
$\as \ud x/x$, where $x$ is the fraction of the longitudinal momentum 
of the incoming hadron and $\as$ is the QCD coupling. This form is 
constant as a function of rapidity, and thus leads naturally to a
rapidity plateau, i.e. a multiplicity $\ud N/\ud y$ approximately independent
of $y$ for scales $\Delta y \lesssim 1/\as$. While at RHIC 
the total collision energy is still so low that the $y$-dependence is 
dominated by large-$x$ physics, at the LHC this plateau has become prominently
visible in the experimental data. At high enough energy there is phase space 
for many gluon emissions in the cascade, each additional one suppressed 
by a factor $\as$ but enhanced by a phase space volume
$\Delta y \sim \ln \sqrt{s}$.
With a factor $1/n!$ from the rapidity ordering
of $n$ gluons one could in fact  very roughly expect
\begin{equation}
\frac{\ud N_g}{\ud y} \sim \sum_n \frac{1}{n!}(\as \ln \sqrt{s})^n \sim \sqrt{s}^{\as} 
\end{equation}
gluons. Again, the experimental energy dependence of multiplicities
in pp ($\sim \sqrt{s}^{0.2}$) and AA ($\sim \sqrt{s}^{0.3}$) collisions
fits in very nicely with this very crude weak coupling, small angle scattering
estimate. This could be contrasted
with a strong coupling picture, which would generically lead to complete baryon 
stopping~\cite{Kovchegov:2009he} and a stronger growth of the multiplicity with 
$\sqrt{s}$.

Eventually the gluons in this cascade will overlap and gluon mergings
will start to compete with splittings. This increase in the gluon density 
corresponds to  the Yang-Mills theory becoming completely
nonperturbative. This happens when the two terms of the
covariant derivative $-iD_\mu = -i\partial_\mu + g A_\mu = p_\mu + g A_\mu$
are of the same order. The (transverse) momentum scale at which this 
happens is referred to as the saturation scale
$\ptt \sim g A_\perp \sim \qs$. When the energy
is high enough, $\qs\gg \lqcd$ and the coupling is still weak.
What results is a picture that has a weak 
coupling $g$ but is still nonperturbative due to the
large gluon field strengths $A_\mu \sim 1/g$. Since the occupation 
numbers of gluonic states $f(k) \sim A_\mu A_\mu\sim 1/\as$ are large,
the gluon field is, to leading order in the coupling, a classical field.

Our understanding of how exactly such a system of strong (but very anisotropic) 
gluon fields thermalizes (or slightly more modestly isotropizes) has
seen significant progress in the recent years (for a review 
see~\cite{Gelis:2015gza}). Pursuing this question in any depth is out of 
scope here, but as a rough outline the process must behave in the 
following way. The initial classical fields, with with occupation numbers
$f(k)\sim \frac{1}{\as} \gg 1$ are very anisotropic due to the longitudinal 
expansion of the system. As gluons start to scatter off the transverse plane
and occupy a larger volume in phase space, the occupation number decreases.
When $f(k) \ll \frac{1}{\as} $ one can switch to a kinetic theory description 
and follow the system towards local equilibrium. Close to isotropy
kinetic theory matches smoothly into a 2$^\mathrm{nd}$ order
viscous hydrodynamical description. A recent weak coupling
calculation using an effective kinetic theory for 
QCD~\cite{Kurkela:2015qoa} arrives, when 
extrapolating to realistic values of $\as$, to times as short
as 1~fm for a matching to viscous hydrodynamics.

\section{Dilute-dense control measurements}

The quintessential experiment for measuring the partonic content of 
a hadron or nucleus is deep inelastic scattering (DIS). 
A good description of the initial state of a heavy ion collision
should be consistent with the 
precise measurements of quarks and gluons  in a proton at HERA 
and with the available electron-nucleus cross section data. It should
also be able to give quantitative predictions for future
measurements at an EIC. This is naturally true for calculations
whose starting point are nuclear pdfs (e.g. 
EPS09~\cite{Eskola:2009uj}). It also applies to 
more recent CGC calculations using e.g. the IPSat or bCGC 
parametrizations~\cite{Rezaeian:2012ji,Rezaeian:2013tka}
and to calculations using the running coupling 
Balitsky-Kovchegov equation (rcBK), e.g.~\cite{Albacete:2010sy,Lappi:2013zma}.
In the classical field CGC picture the initial 
color fields in a heavy collision are calculated in
terms of Wilson lines, whose correlator determines 
the total DIS cross section.

\begin{figure}
\centerline{\includegraphics[width=0.45\textwidth,clip=true,bb = 0 220 355 430]{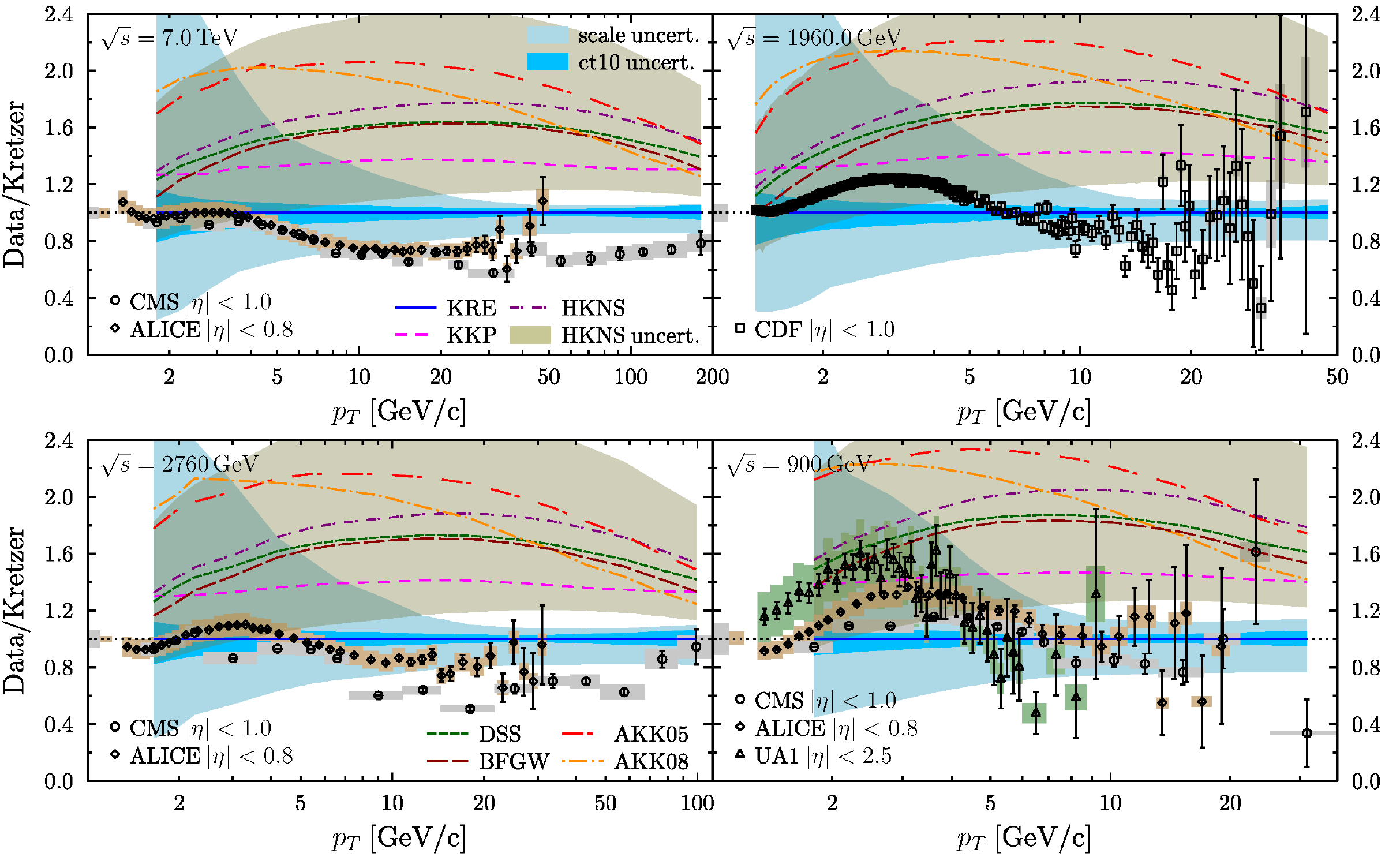}}
\caption{Different fragmentation functions in NLO pQCD compared to CMS data
for charged particle spectra in proton-proton collisions,
from~\cite{d'Enterria:2013vba}.}
\label{fig:ff}
\end{figure}

Another check of QCD descriptions is provided by hadronic
dilute-dense, i.e.  proton-nucleus, collisions.
The most straightforward observables here
are ratios of particle spectra in proton-nucleus collisions to those
in proton-proton ones, normalized with the number of nucleons in the
nucleus, known as $R_{pA}$. At high $\ptt$ these
ratios would be expected to approach unity for many particle species.  This is 
particularly important when moving from minimum bias
proton-nucleus collisions to separate centrality classes, 
where hard prticle production serves as an important consistency 
check of the Glauber modeling~\cite{Adam:2014qja} needed for the
centrality determination. If the normalization of $R_{pA}$
is under control, its small-$\ptt$ behavior is sensitive to the enhancement
of  saturation effects  in a nucleus, especially 
at forward rapidity where one probes smaller
values of~$x$ in the nucleus. 

Typical CGC calculations of hadronic $R_{pA}$ at semihard 
$\ptt$ use two different formalisms. The forward rapidity
``hybrid'' picture starts from  a collinear
quark or gluon from the probe, which is described in terms of 
conventional DGLAP evolved parton distribution functions. This parton 
propagates through the dense color field of the target, picking up an eikonal 
Wilson line.
The other possible treatment is a symmetric $k_T$-factorized description, 
where the produced gluon spectum is obtained as a convolution of 
two unintegrated gluon distributions (ugd's) describing the projectile and 
the target. The ugd is, again, obtained as a Fourier-transform of the
Wilson line correlator probed in DIS. Both formalisms require,
unexceptionally for leading order QCD calculations,  a 
``$K$-factor'' of order $\sim 2$ to fit the spectra, but this normalization 
uncertainty  cancels in the ratio $R_{pA}$.
Unlike in the early days of RHIC,  up to date calculations of 
$R_{pA}$~\cite{Albacete:2010bs,Tribedy:2011aa,Rezaeian:2012ye,Lappi:2013zma} 
use ugd's that have been fit to the  precise HERA data.
In fact, because this data is very constraining, the main quantitative 
difference between these calculations is the treatment of the nuclear geometry
needed to go from a proton to a nuclear target. With $R_{pA}$
predictions differing, e.g. by a factor $\sim 2$ from Ref.~\cite{Albacete:2010bs}
to~\cite{Lappi:2013zma}, this can be a significant
effect.  A similar example of the  importance of a proper treatment
of the nuclear geometry is provided by forward $J/\Psi$ 
production~\cite{Ducloue:2015gfa}.

Even more discriminative power is in principle provided by the absolute
spectra themselves. In a weak coupling QCD calculation the normalization 
has, however, much more uncertainty than the ratio $R_{pA}$.
An additional complication comes from the recent 
observation~\cite{d'Enterria:2013vba} that current fragmentation functions
do not provide a very good description of data 
(see \fig\ref{fig:ff}) 
even in the cases where they should
work well, i.e. proton-proton collisions at higher $\ptt$. This will 
require further development, focusing in particular on the poorly 
constrained gluon fragmentation funtions, see e.g. \cite{deFlorian:2014xna}.
Electroweak probes (photons and
electroweak bosons) are theoretically the cleanest observable probing
weak coupling partonic physics in the nucleus. Here the greater question
is whether the experimental accuracy will be good enough to distinguish 
nuclear effects.

\section{MC Glauber and dynamical models}

Initial matter formation in heavy ion collisions is often parametrized in
terms of a ``Monte Carlo Glauber'' (MCG) model \cite{Miller:2007ri}. One starts
from nucleons distributed inside the nucleus according to a standard 
Woods-Saxon nuclear density profile.  
Nucleons from the colliding projectiles are then deemed to ``collide'' or 
``participate'' if they lie within a short enough distance (determined
by the total inelastic nucleon-nucleon cross section) from each other.
Experimental event-by event distributions of charged particle 
multiplicities or forward calorimeter energies can be roughly modeled
by assuming that the multiplicity or energy is proportional to the number
of ``participant nucleons'' or ``binary nucleon-nucleon collisions''. 
This has led to the MCG model being used both experimentally in 
event centrality classification, and theoretically as a spatial distribution 
of the initial conditions of  hydrodynamical simulations. 

In hydrodynamical calculations microscopical models of particle production are
sometimes contrasted with MCG based on comparisons to 
experimental data. We would like to argue that this is a fundamentally
misleading and unfair comparison. In addition to not being completely
unique (one can e.g. choose the multiplicity or the energy as proportional
to either $\ncoll$ or $\npart$), the MCG model is a purely empirical observation 
without any dynamical mechanism for particle production. It does not,
for example, make any predictions about rapidity or $\sqrt{s}$ dependence.
The MCG model is such a successful description of nuclear geometry that a
similar picture is in fact
built into most modern initial state calculations
which combine it with a microscopical description of particle production
(strings, classical fields, parton scattering \dots).
Thus it is not surprising that, to a first approximation, the centrality 
dependence of bulk observables from these models is close to MCG.
On a conceptual level, if one is interested in understanding microscopical 
QCD dynamics, the meaningful comparison is between different dynamical
models of particle production, not between them and the MCG. One can
ask whether the theory is consistent within itself, or how well the 
(necessarily small) deviations from a 
$\ud N_\mathrm{ch}/\ud \eta \sim \npart$ scaling
that it predicts describe experimental observations.

Let us recall some examples of how a very similar
treatment of the nuclear geometry
is at the heart of many recent initial state calculations.
The MCKLN model~\cite{Drescher:2006ca} explicitly starts
from an MCG calculation that determines, event-by-event, a set of participant
nucleons in each nucleus. One then takes the (now probe-dependent, i.e. 
nonuniversal) saturation scales in the nuclei as proportional to this
$\npart$ and uses them as inputs in a $\ktt$-factorized calculation 
of the initial energy density. The MCrcBK model~\cite{Albacete:2010ad} is 
otherwise similar, but the ad hoc functional form of the ugd is replaced
by a solution of the BK equation. The IPglasma initial state 
model~\cite{Schenke:2012wb} starts from the IPsat~\cite{Kowalski:2003hm} 
parametrization for the dipole scattering amplitude in a nucleus
\begin{equation}
\mathcal{N}_{q\bar{q}} =  
1-e^{ 
-  \frac{\pi^2}{2 \nc}  \as(\mu^2) 
xg(x,\mu^2) 
\sum_{i=1}^A T_p(\bt-\bt_i)
\rt^2 },
\end{equation}
involving fluctuating nucleon positions $\bt_i$ from a Woods-Saxon distribution.
One then determines a local saturation scale $\qs(\bt)$ from this amplitude
and uses it in an MV-model correlator of color charges 
\begin{equation}
\langle \rho^a(\xt) \rho^b(\yt)\rangle \sim \qs^2(\xt)\delta^{ab}
\delta^{(2)}(\xt-\yt),
\end{equation}
from which the classical Yang-Mills (CYM)  field is calculated.
The EKRT model in its present version~\cite{Niemi:2015qia}
uses a fluctuating nuclear thickness profile 
\begin{equation}
T_A(\bt) = \sum_{i=1}^A T_p(\bt-\bt_i) 
\end{equation}
to determine the initial energy density as
\begin{equation} 
\varepsilon(\bt) = \mathcal{F}\left[T_A(\bt)T_B(\bt)\right],
\end{equation}
where the QCD dynamics in the calculation is parametrized by the 
function $\mathcal{F}$. Also string-based models such as
EPOS~\cite{Werner:2010aa} and AMPT~\cite{Lin:2004en} share the
same description of the nuclear geometry.
Since we are still far from being able to calculate nuclear structure from first
principles QCD, at some level the MCG 
geometry has to be put in by hand,
no matter the sophistication of the QCD calculation

Besides the spatial, distribution the calculation of the initial state
should naturally also provide the value of the energy (or entropy) density. 
Ideally this should be done without any adjustable free parameters,
but as we will now discuss this is not fully the case for any of the
currently available calculations.  
In a purely perturbative calculation the initial gluon multiplicity
is not finite, so the question of normalization is intimately tied in 
with the issue of infrared cutoffs. This is a particularly
salient feature of the EKRT model~\cite{Paatelainen:2013eea}, where
the initial parton spectrum calculated in collinear perturbation theory 
has a strong power law dependence on the infrared cutoff. The model introduces
two essential free parameters, the cutoff (saturation scale) $p_0$ and a
normalization coefficient $K_\mathrm{sat}$, which are simultaneously determined
from the final state saturation  criterion 
\begin{equation}
\frac{\ud E_T(\ptt > p_0)}{\ud^2 \xt} = K_\mathrm{sat}\frac{p_0^3 \Delta y}{\pi}
\end{equation}
and by fitting the multiplicity in the most central LHC lead-lead collisions.
In KLN-type initial conditions that use a 
$\ktt$-factorized formula to compute gluon production the procedure is 
slightly different. Because of parton saturation the total gluon multiplicity
is only logarithmically infrared divergent
\begin{equation}
\frac{\ud N}{\ud^2\pt \ud y} =
\frac{C}{\as} \frac{1}{\pt^2} \int_{\kt}  
\varphi_y(\kt) \varphi_y(\pt-\kt),
\end{equation}
where $\varphi_y(\kt)$ is the unintegrated gluon distribution.
Thus, in stead of determining the cutoff together with the normalization,
one usually chooses an infrared cutoff scale (often by restricting
the $\kt$-integral to $\ktt \leq \ptt$; for a different
scheme see~\cite{Tribedy:2011aa}) and then adjusts the normalization
constant $C$ to data. This leaves the gluon mean $\ptt$ (energy per particle)
quite unconstrained and potentially very far from a value extrapolated
backward from the experimental data~\cite{Albacete:2010ad}.

In the CYM calculations, such as the IPglasma 
model~\cite{Schenke:2012wb}, the gluon multiplicity is finite due to nonlinear
effects in the final state~\cite{Blaizot:2010kh} and no infrared 
cutoffs are needed. The Wilson lines $U(\xt)$ that fully determine the initial 
conditions of 
the CYM calculation~\cite{Kovner:1995ja} 
are constrained by DIS data because the  correlator 
\begin{equation}
\frac{1}{\nc} \tr U^\dag(\xt) U(\yt)
\end{equation}
is proportional to the total DIS cross section.
Thus on a conceptual level
there are no free adjustable parameters in the CYM scenario. 
The practical implementation in IPglasma, however, 
uses a procedure that leaves room for small uncertainties. One first
uses a fit to HERA data to get the saturation scale $\qs^2(\bt)$.
This is then input into an MV model calculation 
so that one reproduces the same $\qs$, but not the full 
Wilson line correlator of the DIS fit. Additionally, 
in the MV model in a finite nucleus one must regulate long distance
Coulomb tails of the classical gauge fields with a confinement 
scale parameter. The net result of this procedure is that the Wilson line
in the CYM initial condition does not exactly match the original DIS cross section,
although it is very well constrained by it.

\section{Correlations:  direct signals of the initial state?}
\begin{figure}
\centerline{\includegraphics[width=0.35\textwidth,clip=true]{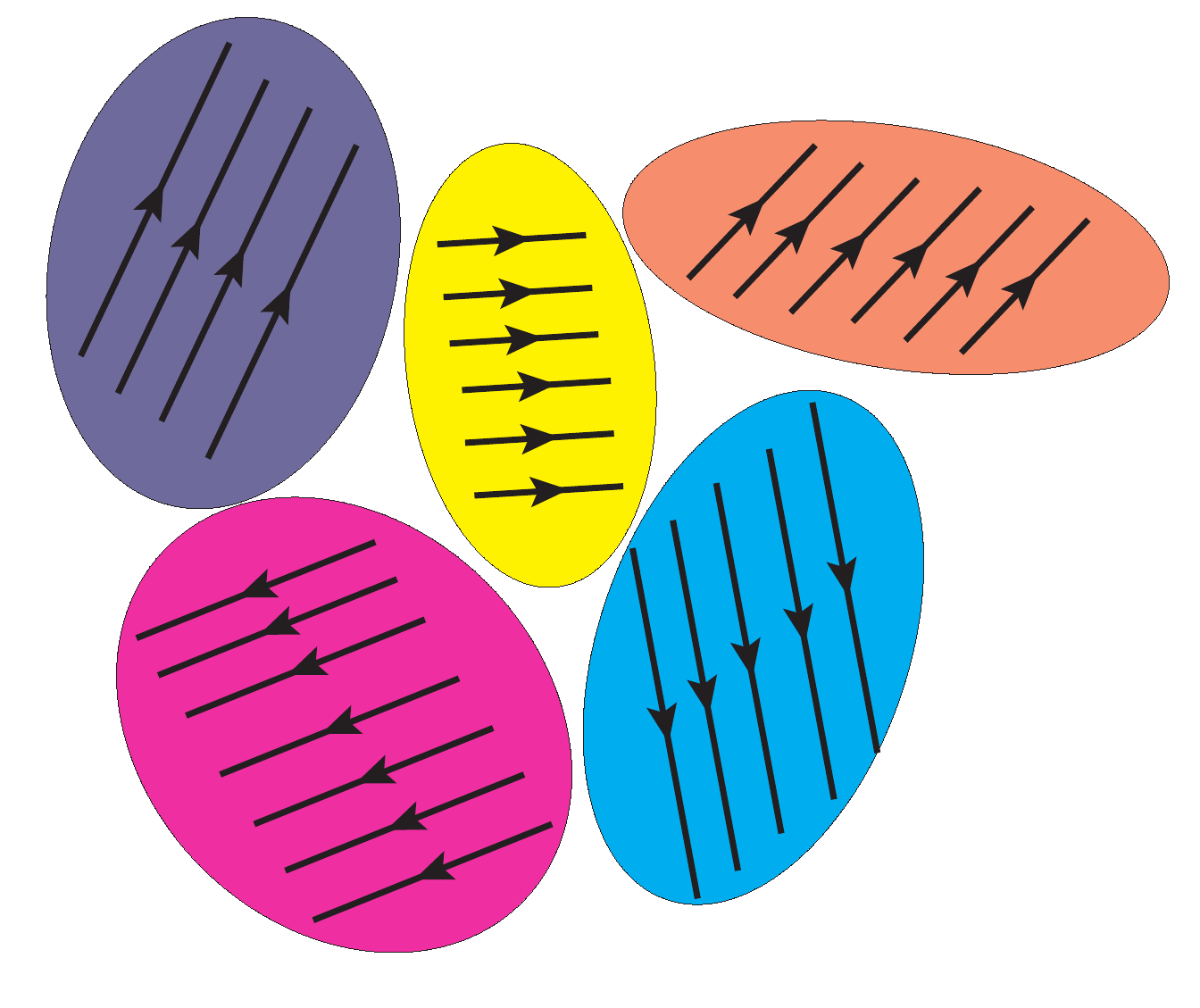}
\begin{tikzpicture}[overlay]
\node[anchor=north west]at (-1.4,1.5) {$1/\qs$};
\draw[line width=2pt,<->](-1.8cm,0.3cm) -- (-0.9cm,2.6cm);
\end{tikzpicture}
}
\caption{Color field domains in the target}
\label{fig:domains}
\end{figure}
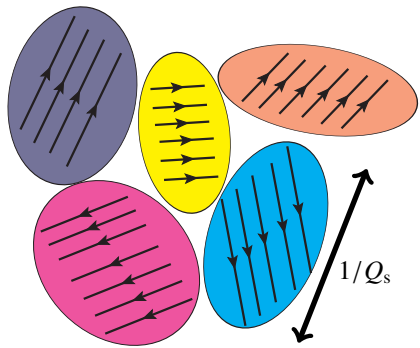

A simple causality argument shows that correlations in particle 
production that have a long range in rapidity 
must have originated in the earliest stage of the collision process.
This is analogous to the way correlations in the cosmic microwave
background are sensitive to  early times, when now separate 
parts of the universe were still in causal contact. Azimuthal flow
coefficients $v_n$ are an important example of long range rapidity 
correlations. They are either explicitly constructed from a 
(usually rapidity separated) multiparticle
correlation function (the ``cumulant method'') or by correlating
particles at one rapidity with a reaction plane determined from particles
in another part of the detector.
The same correlation can be presented in terms of 
a yield per trigger or as a flow coefficient $v_n$; see e.g.~\cite{Aad:2014lta}.
 The most important origin of these 
correlations, especially in large collision systems, is
the geometry of the initial state: the distribution 
of matter in the transverse plane is by definition felt at all rapidities.
Geometrical correlations are, however, in position space, and strong
enough collective final state interactions  are needed to transform
them into observable momentum space distributions.

It has more  recently~\cite{Dumitru:2010iy,Kovner:2010xk} been pointed out that,
in particular in small collision
systems, the domain structure in the target color field can also 
generate azimuthal correlations directly in momentum space
without  final state collective phenomena. These correlations follow in a very
intuitive way from the CGC picture. The usual picture of particle 
production in this case is that of an individual quark or gluon 
(whose number is given by a conventional pdf) passing through the 
strong color field of the target and being deflected to some transverse momentum.
Since the target color field consists of domains of size $\sim 1/\qs^2$
(see \fig\ref{fig:domains}), incoming
particles in the same color state and hitting the same domain experience a
similar deflecting color field and become correlated. The mechanism
generates correlations that are suppressed 
by the number of independent domains and colors as
 $\sim 1/(\nc^2 \qs^2 S_\perp)$, 
where $S_\perp$ is the size of interaction
area. Thus, in contrast to collective flow effects, these correlations
are enhanced in small collision systems.

These correlations have recently been analyzed by different authors in calculations
that share the same physical picture, but differ in the approximations used to calculate
the correlations of the target color fields. The ``Glasma graph'' 
calculations~\cite{Dumitru:2010iy,Dusling:2012iga}
linearize in the color charge density of the target and assume Gaussian correlations
between the domains. In the ``color field domain model''~\cite{Dumitru:2014dra}
one also linearizes in the target field, but adds an additional non-Gaussian correlation.
On the other hand one can perform a full nonlinear calculation in the dilute-dense 
limit~\cite{Dumitru:2014vka,Lappi:2015vha} without adding non-Gaussian correlations. 
By performing a full CYM simulation~\cite{Schenke:2015aqa}
one can also include pre-equilibrium collective
effects in the calculation of the azimuthal dependence.
For a more detailed reecnt discussion of the differences and similarities 
see~\cite{Lappi:2015vta}. So far all of these calculations assume 
parametrically small rapidity separations $\Delta y\lesssim 1/\as$, but there have
been proposals to extend the calculation also to parametrically
large rapidity intervals~\cite{Iancu:2013uva}.

\section{Conclusions}

In conclusion, recent years have seen a huge progress towards a more well defined
weak coupling picture of the initial stages of a heavy ion collision. 
Studies of thermalization are moving from qualitative to quantitative, and the importance of 
interfacing with  kinetic theory for the later stage of equilibration is 
becoming understood.  One wants the initial stage calculation to be consistent 
with perturbative probes and  control measurements in dilute-dense collisions. We emphasized
that the MC Glauber-like geometry is built into the more modern dynamical 
models of the initial state. 
Finally we discussed the azimuthal structure of long range rapidity correlations, 
i.e. the $v_n$-coefficients, especially in small systems, where the interplay between initial 
and final state collective effects remains to be quantitatively sorted out.

 This work has been supported by the Academy of Finland, projects 
267321 and 273464.

\bibliographystyle{elsarticle-num}
\bibliography{spires}

\end{document}